\documentclass[prl,aps,floatfix,,twocolumn,groupedaddress,amsmath,showpacs,twoside]{revtex4}
\usepackage{graphicx}
\usepackage{chemarr}
\usepackage{times}
\usepackage{hyperref}
\usepackage{version}
\usepackage{color}

\newcommand{\fig}{Fig.}
\newcommand{\eq}{Eq.}

\newcommand{\p}{\partial}
\newcommand{\ee}{\textrm{e}}

\newcommand{\on}{\textrm{on}}
\newcommand{\off}{\textrm{off}}
\newcommand{\kon}{k^{\textrm{on}}}
\newcommand{\koff}{k^{\textrm{off}}}
\newcommand{\pon}{p_{\on}}
\newcommand{\poff}{p_{\off}}
\newcommand{\gon}{G_{\on}}
\newcommand{\goff}{G_{\off}}
\newcommand{\hon}{h_{\on}}

\newcommand{\Ht}{{\tilde{h}}}
\newcommand{\sce}{\textrm{SCE}}
\newcommand{\sse}{\textrm{SSE}}

\providecommand{\avg}[1]{\left \langle #1 \right \rangle}

\providecommand{\addhyphen}[1]{#1.---}

\makeatletter
\renewcommand \paragraph{%
  \@startsection
    {paragraph}%
    {4}%
    {\parindent}%
    {\z@}%
    {-.1em}%
    {\normalfont\normalsize\itshape\addhyphen}%
}%
\makeatother

\begin{document}

\title{Exact solution of a model DNA--inversion genetic switch with
orientational control}

\author{Paolo Visco}

\affiliation{SUPA, School of Physics, The University of Edinburgh,
Mayfield Road,
Edinburgh EH9 3JZ, UK}

\author{Rosalind J. Allen}

\affiliation{SUPA, School of Physics, The University of Edinburgh,
Mayfield Road,
Edinburgh EH9 3JZ, UK}

\author{Martin R. Evans}

\affiliation{SUPA, School of Physics, The University of Edinburgh,
Mayfield Road,
Edinburgh EH9 3JZ, UK}

\date{\today}

\begin{abstract}
DNA inversion is an important mechanism by which bacteria and
bacteriophage switch reversibly between phenotypic states. In such
switches, the orientation of a short DNA element is flipped by a
site-specific recombinase enzyme. We propose a simple model for a DNA
inversion switch in which recombinase production is dependent on the
switch state (orientational control). Our model is inspired by the
{\em{fim}} switch in {\em{Escherichia coli}}. We present an exact
analytical solution of the chemical master equation for the model
switch, as well as stochastic simulations. Orientational control
causes the switch to deviate from Poissonian behaviour: the
distribution of times in the on state shows a peak and successive flip
times are correlated.
\end{abstract}

\pacs{87.18.Cf,87.16.Yc,82.39.-k}

\maketitle

Reversible and heritable stochastic switching between two different
states of gene expression is a common phenomenon among bacteria and
bacteriophage, known as phase variation. Phase variation is often
linked to pathogenesis, and may help bacteria survive fluctuating
environmental conditions (e.g., a host immune
system)~\cite{woude2004}. An important molecular mechanism for phase
variation is site-specific DNA inversion~\cite{woude2006}. Here, a
short piece of DNA (the ``invertible element'') is excised from the
genome and reinserted (strand-by strand) in the opposite orientation
by a site-specific recombinase enzyme binding to sequences at the ends
of the invertible element. Different states of gene expression
correspond to the two orientations of the invertible element (``switch
states''). A well-known example is the {\em{fim}} genetic regulatory
system, which controls the production of type 1 fimbriae in {\em
Escherichia coli}. These fimbriae are important in
uropathogenesis~\cite{blomfield2001}. In the {\em{fim}} system, the
FimE recombinase is produced more strongly when the switch is in the
on state than in the off state~\cite{kulasekara1999,joyce2002}. This
phenomenon is known as orientational control.

In this Letter, we present a simple and general stochastic model for a
DNA inversion switch with orientational control. We solve this model
analytically, allowing us to determine the range of stochastic
switching behaviour possible for this type of switch. We find that
non--Poissonian behaviour occurs, resulting in a peak in the
probability distribution of time spent in the on state and
correlations between successive flips. Such non--Poissonian behaviour
could have important effects on the population dynamics of switching
microbes in changing environments. One key parameter (the
concentration of the recombinase not under orientational control)
controls the degree to which our model is non--Poissonian; this
parameter corresponds to the main point of environmental regulation
for the {\em{fim}} switch. In contrast, bistable genetic switches such
as positive feedback loops~\cite{dubnau2006} or mutually
repressing genes~\cite{cherry2000,warren2005}
in general show only Poissonian behaviour (exponential waiting time
distributions and uncorrelated flips).

\begin{figure} 
\includegraphics[width=0.85\columnwidth,clip=true]{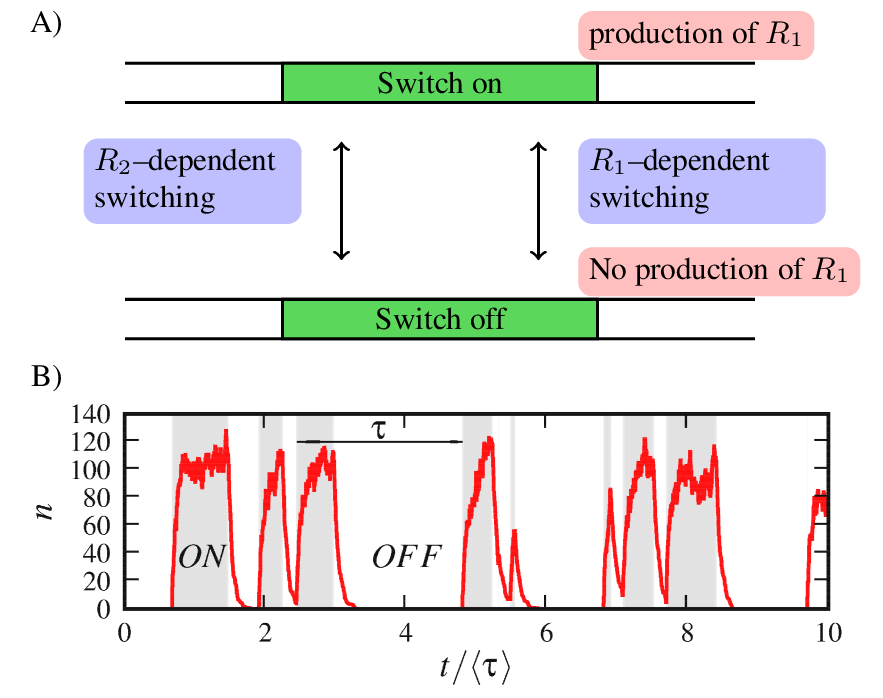}
\caption{\label{fig:scheme} (color online) A) Schematic illustration
of the model. B) A typical simulation trajectory. The solid line
represents the time evolution of the number $n$ of $R_1$--molecules,
the shading denotes the switch position and $\tau$ indicates the
duration of a period of the switch. Parameter values are $k_1=1$,
$k_2=100$, $k_3^{\on}=0.001$, $k_3^{\off}=0$ $k_4^{\on}=0.1$ and
$k_4^{\off}=0.1$.}
\end{figure}

\paragraph{The Model} 
Our model DNA inversion switch, illustrated in \fig~\ref{fig:scheme},
contains three elements: the invertible DNA element and two types of
recombinase enzyme ($R_1$ and $R_2$). The invertible element has two
possible orientations (the ``on'' and ``off'' states). These
correspond to alternative patterns of gene expression, leading to
different phenotypic states; however, we model here only the core of
the switch and not its downstream effects. The switch can be flipped
between its two orientations by either of the recombinases. The
concentration of recombinase $R_2$ is assumed to be fixed, while the
production of $R_1$ depends on the switch state: $R_1$ is produced
only in the on state. This feature of the model constitutes its
orientational control and leads to its non--Poissonian behaviour. The
model is represented by the following reaction scheme:
\begin{subequations}
\label{eq:react}
\begin{align}
\label{eq:reacta}
R_1 & \stackrel{k_1}{\longrightarrow}\emptyset & S_{\on} &
\stackrel{k_2}{\longrightarrow} S_{\on}+R_1 \\
\label{eq:reactb}
S_{\on} + R_1 & \xrightleftharpoons[\koff_3]{\kon_3} S_{\off} + R_1 & 
S_{\on} & \xrightleftharpoons[\koff_4]{\kon_4} S_{\off} \,\,.
\end{align}
\end{subequations}
Here, $S_{\on}$ and $S_{\off}$ denote the on and off states of the
switch. Reactions (\ref{eq:reacta}) describe the production and decay
of recombinase $R_1$: $R_1$ is removed from the system with rate
constant $k_1$ (due to cell growth and division, which we do not model
explicitly), and is produced at rate $k_2$ only when the switch is in
the on state. Reactions (\ref{eq:reactb}) describe switch
flipping. This may be catalysed by $R_1$ with rate constants $\kon_3$
($\on$ to $\off$) and $\koff_3$ ($\off$ to $\on$). We shall mainly
consider here the case $\koff_3=0$. Recombinase $R_2$ can also
catalyse switch flipping. The concentration of $R_2$ (fixed in our
model) is not explicitly included in the reaction scheme, but is
implicit through a dependence of the rate constants $k_4^{\on/\off}$
on the $R_2$ concentration.

We note that mean-field, macroscopic rate equations corresponding to
the above reaction scheme yield only one steady-state solution
corresponding to the average switch state and average concentration of
$R_1$. The underlying deterministic structure of the model is thus not
bistable. In this sense, our model is fundamentally different from the
bistable reaction networks presented in
\cite{dubnau2006,cherry2000,warren2005}.

Our model is inspired by the {\em{fim}} genetic regulatory
system~\cite{blomfield2001}. In analogy with {\em{fim}}, $R_1$ in our
model represents FimE while $R_2$ represents FimB. Environmental
stimuli such as nutrient conditions and temperature act on the fim
switch largely through changes in the level of FimB
\cite{wolf2002,chu2007}; this would correspond to variation of our key
parameters $k_4^{\on/\off}$.  However, our model is highly simplified
in comparison to {\em{fim}}, in that it neglects cooperative and
competitive recombinase binding and the effects of other DNA binding
proteins.  Our objective in this work is not to model the details of
the {\em{fim}} system, as other authors have
done~\cite{wolf2002,chu2007}, but rather to address general questions
about the behaviour of this type of switch. Our model is designed to
be as simple as possible while retaining the key features of DNA
inversion and orientational control; its simplicity allows us to
obtain analytical results and to explore a wide range of parameter
space.

\begin{figure}
\includegraphics[width=0.95\columnwidth,clip=true]{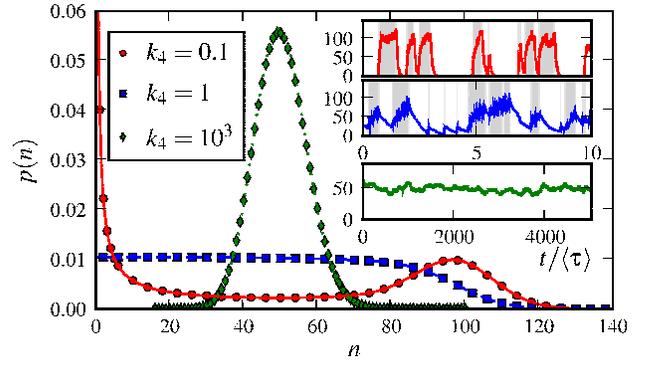}
\caption{\label{fig:ponpoff} (color online) Probability distribution
  for the number $n$ of $R_1$ molecules, with
  $k_4^{\on}=k_4^{\off}=k_4$, for different values of $k_4$. The
  symbols show simulation results and the solid lines are the
  theoretical predictions. For the case where $k_4=10^3$, the (dotted)
  line is a Poisson distribution with parameter $(1+
  k_4^{\on}/k_4^{\off})^{-1}k_2^{\on}/k_1$. The other parameters are:
  $k_1=1$, $k_2^{\on}=100$, $k_3^{\on}=0.001$ and $k_3^{\off}=0$. The
  insets show the typical simulation trajectories for each
  distribution, where the time is expressed in units of the mean
  period $\avg{\tau}$.}
\end{figure}

 We simulated the reaction scheme (\ref{eq:react}) using a continuous
time Monte Carlo scheme~\cite{gillespie1976}. A typical trajectory is
shown in \fig~\ref{fig:scheme}, where we plot the number $n$ of $R_1$
molecules and the switch state as functions of time. When the switch
is in the on state, $n$ increases (on average) towards a plateau value
of $k_2/k_1$, while in the off state $n$ decays towards zero.  We now
obtain an exact analytical solution for the case where
$\koff_3=0$. This case is relevant to the {\em{fim}} switch, where
FimE catalyses almost exclusively on to off switching. In the
following, all our analytical results will correspond to $\koff_3=0$,
while simulation results will be presented also for
$\koff_3>0$. Analytical results for $\koff_3>0$ will be published
elsewhere.
\paragraph{Steady State} 
We first consider the statistics of $n$ in the steady state and
compute the long time, joint probability $p_s(n)$ that the switch is
in state $s$ and there are $n$ molecules of $R_1$. The system of
birth--death equations for $\pon(n)$ and $\poff(n)$ becomes in the
steady state
\begin{subequations}
\label{eq:masterp}
\begin{multline}
\label{eq:masterpon}
(n+1) k_1 \pon(n+1)+ k_2 \pon(n-1) + \koff_4 \poff(n)\\ = (n k_1 +
 k_2 + n \kon_3 + \kon_4) \pon(n) \,\,,
\end{multline}
\vspace*{-0.7cm}
\begin{multline}
\label{eq:masterpoff}
(n+1) k_1 \poff(n+1) + n \kon_3 \pon(n) + \kon_4 \pon(n)\\ = (n
k_1 + \koff_4) \poff(n)\,\,\,.
\end{multline}
\end{subequations}
In order to decouple the above set of equations, we solve
(\ref{eq:masterpon}) for $\poff$, then insert the result into
(\ref{eq:masterpoff}) to give a decoupled equation for
$\pon$. Introducing the generating function $G_s(z)=\sum_{n}z^n
p_s(n)$ (where $s=\{\on,\off\}$), the decoupled equation for $\pon$
reduces to a second order differential equation for $\gon$. Defining
then a new variable $u \equiv u_z= k_2 z /(k_1 + \kon_3) - k_1 k_2 /
(k_1 + \kon_3)^2$, the latter equation reads:
\begin{equation}
u \gon''(u)+ (a-u) \gon'(u) - b \gon(u)=0\,\,,
\end{equation}
where $a=1+u_1+(\kon_4 + \koff_4)/(k_1 + \kon_3)$ and
$b=1+\koff_4/k_1$. Expanding the solution as a regular power series
(i.e. $\gon(u)= \sum_m a_m u^m$), one finds that $\gon(z)=a_0 \, {_1
F_1}(b,a,u)$, where $_1F_1$ is a confluent hypergeometric function and
$a_0$ is an integration constant which can be determined with the
normalisation condition $\sum_n \pon(n) + \poff(n)=1$ (or equivalently
$\gon(1)+\goff(1)=1$). This result can be rewritten in terms of the
original variable $n$ as
\begin{equation}
\label{eq:ponsol}
\pon(n)=a_0 \frac{(u_1-u_0)^n}{n!} \frac{(b)_n}{(a)_n}
{_1F_1}(b+n,a+n,u_0)\,\,,
\end{equation}
where $(\alpha)_n=\alpha (\alpha+1) \dots (\alpha+n-1)$.  An
expression for $\poff$ can be derived by inserting \eq
(\ref{eq:ponsol}) into Eq. (\ref{eq:masterpon}). 
In \fig \ref{fig:ponpoff} we compare the result for 
$p(n) = \pon +\poff$ to simulations, obtaining
perfect agreement.
Figure~\ref{fig:ponpoff} also illustrates the effects of the different
timescales for switch flipping and production/decay of $R_1$. For
small $k_4$, switch flipping is slow compared to the rate of change of
$n$. In this case, when the switch is in the on state, the number of
recombinase has time to reach a plateau before the switch flips
off. The on and off switch states are then each associated with a
different value of $n$ and the distribution $p(n)$ is bimodal. In
contrast, when $k_4$ is large, the switch flips back and forth much
more rapidly than recombinase production or removal. Then, the
fraction of time spent in the on state is
$k_4^{\off}/(k_4^{\off}+k_4^{\on})$, and $p(n)$ tends to the Poisson
distribution expected for a birth-death process with birth rate
$k_2^{\on}k_4^{\off}/(k_4^{\off}+k_4^{\on})$ and death rate $k_1$.

\paragraph{Flipping time distributions}
To determine how orientational control affects switch function, we
compute flipping time distributions. The flipping time $T$ can be
defined in two different ways. In the first scenario, which we call
the {\em{Switch Change Ensemble}} ($\sce$), we define $T$ as the time
spent in a particular switch state---for example, $F_{\on}^{\sce}(T)$
is the probability distribution for the time between the moment the
switch enters the on state and the moment it flips from the on to the
off state. In the second scenario, which we call the {\em{Steady State
Ensemble}} ($\sse$), we start observing the cell at a random moment
and measure the time interval between this moment and its next flip
into the other state. $F_{\on}^{\sse}(T)$ and $F_{\off}^{\sse}(T)$ may
be relevant to the response of a population of switching cells to a
sudden environmental change. They also correspond to an experiment
where one measures the time until the next flip, for cells sampled in
the steady state \cite{footnote}.  To compute these
distributions, we define $F_s(T|n_0)$ as the probability that the
system begins at $t=0$ in the $s$ state with $n_0$ recombinase and
flips for the first time at $T$.  Note that for $\koff_3=0$, the off
to on flipping process does not depend on $R_1$ and is governed by
$k_4^{\off}$; thus $F_{\off}(T) = \koff_4 \exp( -\koff_4 T)$ is
independent of $n_0$ (for both the $\sce$ and the $\sse$). However,
the on to off flipping rate is $n_0$ dependent, so that we average
over the ensemble of initial states (characterised by the probability
$W_\on(n)$ of having $n$ recombinase at the start of our measurement)
to obtain the flip time distribution $F_\on(T) = \sum_{n_0} W_\on(n_0)
F_\on(T|n_0)$.  For the $\sce$, the initial condition is taken just
after a flip, which implies that $W_{\on}^\sce(n_0) =
\poff(n_0)/G_{\off}(1)$. For the $\sse$, the initial condition is
sampled in the steady state, yielding $W_\on^\sse (n_0) =
p_\on(n_0)/G_{\on}(1)$.  To compute $F_\on(T)$, we first define the
survival probability $h_\on(n,t)$ that, at time $t$, the switch is in
the on state with $n$ $R_1$ molecules, without having flipped, given
the initial condition $h_\on(n_0,0)= W_\on(n_0)$. The evolution
equation for $h_\on$ is:
\begin{multline}
\label{eq:survon}
\p_t \hon(n,t) = (n+1) k_1 \hon(n+1,t) + k_2 \hon(n-1,t) \\ - (n
k_1 + k_2 + n \kon_3 + \kon_4) \hon(n,t) \;.
\end{multline}
Defining a generating function $\Ht_\on(z,t)= \sum_n z^n h_\on(n,t)$,
it follows that $F_\on(t)=-\p_t \Ht_\on(1,t)$. \eq (\ref{eq:survon})
reduces to a partial differential equation for $\Ht_\on(z,t)$ which
has the initial condition $\Ht_\on(z,0)=\sum_{n_0} W_\on(n_0)
z^{n_0}$. Its solution is
\begin{multline}
\label{eq:hton}
\Ht_{\on}(z,t)=  \ee^{-t (\kon_4+ k_2 (1- k_1 \tau_{\on}))}
\ee^{k_2 \tau_{\on} (z-k_1 \tau_{\on}) (1- \ee^{-t/\tau_{\on}})} \\
\times
\Ht_\on(k_1 \tau_{\on} + \ee^{-t/\tau_{\on}} (z-k_1 \tau),0) \,\,\,,
\end{multline}
where $\tau_{\on}= (k_1 + \kon_3)^{-1}$.  $F_\on(T)$ can  then be
computed for the different measurement scenarios using $\Ht_\on(x,0) =
G_s(x)/G_s(1)$ with $s= {\rm off}$ for  $\sce$ and $s =
{\rm on}$ for  $\sse$.

\begin{figure}
\label{phase_diagr}
 \includegraphics[width=0.875\columnwidth,clip=true]{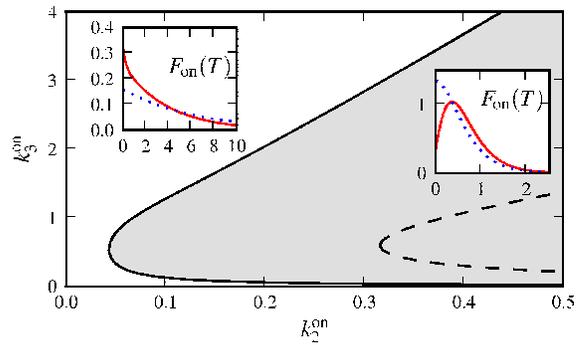}
 \caption{\label{fig:ph_d} (color online) Occurrence of a peak in the
 flipping time distribution $F_\on^\sce(T)$, for $k_4=0.1$. In the
 shaded region the peak is present, and it vanishes outside this
 region. The dashed line shows the same result for $k_4=0.25$. As
 $k_4$ increases, the range of parameters for which there is a peak
 decreases. The unit of time is set by $k_1$ (i.e. $k_1=1$). The
 insets show examples of $F_\on^\sce(T)$ (solid lines) and
 $F_\on^\sse(T)$ (dotted lines). In the left inset, for
 $(k_2,\kon_3,k_4)=(0.2,4,0.1)$, $F_\on^\sce(T)$ is peaked, while in
 the right inset, for $(5,1,0.1)$, $F_\on^\sce(T)$ shows no peak.}
\end{figure}

\paragraph{Peak in the distribution} 
Our results, illustrated in \fig \ref{fig:ph_d}, show a striking
effect of orientational control for this model switch: for the $\sce$,
we can obtain a peak in the flipping time distribution. Such a peak
has been postulated for the {\em{fim}} switch~\cite{wolf2002}, where
it might imply that the switch tends not to leave the on state before
it has had time to synthesise
fimbriae~\cite{wolf2002,chu2007,blomfield2001}. Time spent in the on
state may also influence recognition by the host immune system. This
peak in $F_{\on}^\sce$ is a consequence of the feedback between the
switch state and the level of $R_1$. Once in the on state, the rate
of on to off flipping increases with time as $R_1$ is produced.  In
contrast, for a Poissonian switch, the rate of flipping is constant in
time.  For a peak to occur, the slope of $F_{\on}^\sce$ at the origin
must be positive, which implies
\begin{equation}
\label{eq:ineq}
k_2  - (\kon_4)^2/\kon_3 -  (k_1 + 2 \kon_4) \avg{n_0} 
- \kon_3 \avg{n_0^2} >0 ,
\end{equation}
where $\avg{\dots}$ denotes an average using the weight $W_\on$. The
l.h.s. of (\ref{eq:ineq}) can be evaluated numerically using the exact
result (\ref{eq:ponsol}) to compute $\langle n_0\rangle$ and $\langle
n^2_0\rangle$.  We can then determine the regions of parameter space
where a peak exists, as shown in the shaded region of
\fig~\ref{fig:ph_d} for the SCE. Our results show that the presence of
a peak is favoured by large values of $k_2$ (strong production of
$R_1$ in the on state), and suppressed by very large values of
$\kon_3$ (strong $R_1$--mediated on to off switching) or by very
small values of $\kon_3$ (switching dominated by $R_1$--independent
mechanism). Likewise, when $k_4$ is increased, the range of values
over which the peak exists is decreased, since $R_1$--independent
Poissonian switching tends to dominate. For the $\sse$, on the other
hand, we did not find any parameter values where inequality
(\ref{eq:ineq}) is verified. The peak in the SCE appears because the
number of recombinase, and hence the flipping probability, is
typically low immediately after entering the on state and increases
significantly thereafter. In contrast, in the SSE one typically starts a
measurement when $n$, and hence the flipping probability, is already
high. This tends to suppress the peak in the SSE flipping time
distribution.

\begin{figure}
 \includegraphics[width=0.875\columnwidth,clip=true]{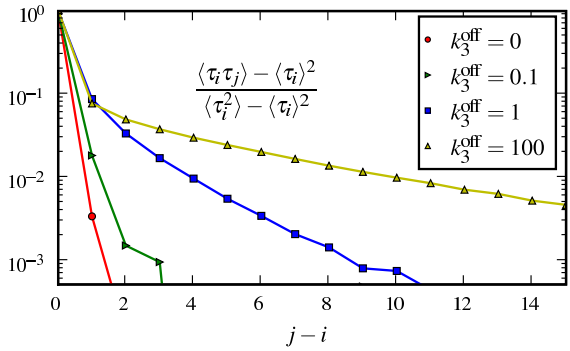}
 \caption{\label{flip} (color online) Correlation function
$\left[\langle \tau_i \tau_j \rangle - \langle \tau_i \rangle ^2
\right]/\left[\langle \tau_i^2 \rangle - \langle \tau_i \rangle ^2
\right]$ between switch periods $\tau_i$ and $\tau_j$, for several
values of the rate $k_3^{\off}$. These are simulation results for
$k_1=1$, $k_2^{\on}=5$, $k_3^{\on}=1$, $k_4^{\on}=k_4^{\off}=0.1$.}
\end{figure}

\paragraph{Correlated flips}
Another potentially important effect of the feedback between the
switch state and the production of recombinase $R_1$ may be to cause
correlations in the waiting times between successive flips (for
example, a particularly short time before a flip might lead to
subsequent flips occurring in quick succession). Such correlations
might allow a population of switching microbes to ``remember'' the
history of past environmental changes. We define the {\em switch
period} $\tau_i$ as the time from when the switch enters the on state
from the off state for the $i$th time, until it enters the on state
for the $(i+1)$th time [cf. \fig~\ref{fig:scheme}]). In
\fig~\ref{flip}, we plot simulation results for the correlation
function for switch periods $\tau_i$ and $\tau_j$ , as a function of
$j-i$. For $\koff_3 = 0$, weak correlation is observed between
subsequent periods $\tau_i$ and $\tau_{i+1}$. Correlations are weak
because when $\koff_3 = 0$, the off to on switching process does not
depend on $R_1$ and is an uncorrelated Poisson process. When $\koff_3
\ne 0$, correlations are much stronger and extended correlated
sequences of flips emerge.

\paragraph{Discussion}
We have presented a generic model for a DNA inversion switch with
orientational control. By solving the model analytically in the case
$k_3^{\off}=0$ (relevant to the {\em{fim}} switch), and using
stochastic simulations, we have shown that this type of switch can
display markedly non--Poissonian behaviour, including a peaked
flipping time distribution for intermediate values of $\kon_3$ and,
for $k_3^{\off} > 0$, correlated sequences of flips. Non--Poissonian
behaviour has been postulated to be a consequence of orientational
control~\cite{blomfield2001,wolf2002}. The model presented here allows
us to analyse the origins and effects of this behaviour in detail, and
provides analytical results which can be used as a basis for more
complex models~\cite{holden2004}. Suggested evolutionary roles for
orientational control include rapid response to environmental
change~\cite{chu2007}, as well as peaked flipping time distributions
\cite{blomfield2001,wolf2002}. This study raises interesting questions
about the consequences of non--Poissonian switching for population
dynamics in changing environments. Several models have been proposed
for the growth of populations of switching cells in stochastically and
periodically changing environments (see, for example
\cite{thattai}). These models assume Poissonian switch flipping. The
analytical solutions presented here should make it possible to extend
such models to the case of non-Poissonian flips.  Non-Poissonian
switch flipping opens up the possibility that lineages of cells may
`remember' (in a statistical sense) the history of their recent
phenotypic states. This is likely to have important consequences for
models which include selection according to the fitness of different
phenotypic states in changing environments.  We speculate that
bacteria which adopt a non-Poissonian flipping strategy may be able to
maximise the evolutionary advantages of using some knowledge of the
likely future behaviour of the environment, combined with the benefits
of a stochastic strategy as an insurance against sudden and
unpredictable environmental changes. These avenues will be the subject
of future work.

\begin{acknowledgments}
The authors are grateful to A. Adiciptaningrum, G. Blakely, M. Cates,
C. Dorman, A. Free, D. Gally, N. Holden, S. T{\u{a}}nase-Nicola and
S. Tans for valuable discussions. R.J.A. was funded by the RSE.
This work was funded by the EPSRC under grant EP/E030173.
\end{acknowledgments}


\begin{thebibliography}{19}
\bibitem{footnote} {Experiments are likely to be complicated by
division events between flips (not considered here).}
\end{thebibliography}
\end{document}